\documentclass{aa}
\usepackage{graphics,color}
\usepackage{txfonts}
\begin{document}
   \title{Long period variables in 47\,Tuc: direct evidence for lost mass}

   \author{T. Lebzelter
          \inst{1}
          \and
          P.R. Wood\inst{2}
          }

   \offprints{T. Lebzelter}

   \institute{Department for Astronomy (IfA), University of Vienna,
              T\"urkenschanzstrasse 17, A-1180 Vienna, Austria\\
              \email{lebzelter@astro.univie.ac.at}
			\and Research School for Astronomy \& Astrophysics,
			Australian National University, Cotter Road, Weston Creek ACT 2611, Australia}

   \date{Received ; accepted }

   \abstract{
We have identified 22 new variable red giants in 47\,Tuc and determined periods
for another 8 previously known variables.  All red giants redder than
$V$-$I_{\rm c}$\,=\,1.8 are variable at the limits of our detection threshold,
which corresponds to $\delta V$\,$\approx$\,0.1 mag.  This colour limit
corresponds to a luminosity log\,$L$/L$_{\odot}$=3.15 and it is considerably
below the tip of the RGB at log\,$L$/L$_{\odot}$=3.35.  Linear non-adiabatic
models without mass loss on the giant branch can not reproduce the observed PL
laws for the low amplitude pulsators.  Models that have undergone mass loss do
reproduce the observed PL relations and they show that mass loss of the order
of 0.3 M$_{\odot}$ occurs along the RGB and AGB.  The linear pulsation periods do not
agree well with the observed periods of the large amplitude Mira variables,
which pulsate in the fundamental mode.  The solution to this problem appears to
be that the nonlinear pulsation periods in these low mass stars are
considerably shorter than the linear pulsation periods due to a rearrangement
of stellar structure caused by the pulsation.  Both observations and theory
show that stars evolve up the RGB and first part of the AGB pulsating in low
order overtone modes, then switch to fundamental mode at high luminosities.

   }

   \maketitle

\section{Introduction}

It is now well established that pulsating red giant stars lie on a
series of up to six parallel period-luminosity (PL) sequences (e.g. Wood et al. \cite{wm99}, Wood \cite{wood00};
Ita et al. \cite{ita04}; Soszy\'{n}ski et al. \cite{sos04}; Fraser et al. \cite{fra05}).
The stars known to populate these sequences are generally field variables
so that their metallicities, masses and ages are not known individually, although the luminosities are
well known because the stars lie in stellar systems at a known distance, such as the SMC and LMC.  The dispersion in
the PL relations is much larger than the amplitude of the pulsation, especially for the small amplitude variables,
and this dispersion is almost certainly a consequence of the dispersion in mass and metallicity at a given pulsation period.

Comparison of the PL relations with theory (Wood et al. \cite{wm99}) shows that four of
the PL sequences (sequences A, B and C of Wood et al. \cite{wm99}, or sequences
A, B, C and C$^{\prime}$ of Ita et al. \cite{ita04}) can be broadly explained by radial pulsation in the
lower order modes.  However,
a real test of the theoretical models requires comparison with stars of known
metallicity, age and (initial) mass since the fits of theory to data in Wood et al. (\cite{wm99})
and Ita et al. (\cite{ita04}) are not particularly good.

Globular clusters are well suited for carrying out such a comparison, and for studying the relation 
between pulsation, mass loss and stellar evolution along the giant branch, since
fundamental parameters like initial mass, luminosity and metallicity are well known.
However, only a small number of pulsating red giants, hereinafter referred to as long-period variables or LPVs, 
is known in any given globular
cluster (Clement et al.\,\cite{clement01}), with 47 Tuc having the most known LPVs (14).  This is an insufficient
number of variables for clearly defining PL sequences, especially as many of the periods
of the LPVs are poorly determined.  The origin of this problem is that most search programs for variable stars in
globular clusters were designed to optimize detection and period
determination for variables with periods of about one day or less. LPVs
with their periods of a few ten to a few hundred days have
therefore not been well surveyed and the existing samples are far
from complete.

To rectify this situation, we have started a search program for long period variables in
Galactic Globular clusters (Lebzelter et al. \cite{lebz04}).  The relatively large number of already known or proposed long period
variables in 47\,Tuc made it a good starting point.  Here, we present
the results for 47\,Tuc (NGC 104), the first cluster analyzed.  We adopt the following 
properties for 47 Tuc: $(m-M)_{V}$=13.5$\pm$0.08 (Gratton et al. \cite{gratton03}); 
metallicity [Fe/H]$=$$-$0.66 (Carretta \& Gratton \cite{cg97}); 
interstellar reddening $E(B-V)$$=$0.024; an age of 11.2$\pm$1.1\,Gyr 
(Gratton et al.\,\cite{gratton03}); and a turnoff mass between
0.86 and 0.9\,$M_{\sun}$.  In a recent paper discussing radial velocity variations
in known LPVs in 47 Tuc (Lebzelter et al.\,\cite{LWHJF05}) we gave
more discussion of the properties of 47 Tuc and its variables and we refer the reader to
that paper for details.

\section{Observations}

\subsection{Mount Stromlo data}
Our monitoring program of 47\,Tuc started in August 2002 at the 50\,inch telescope
at Mount Stromlo. This telescope was equipped with a two channel camera 
used earlier for the MACHO experiment (Alcock et al.\,\cite{macho92}). 
The camera obtained two images in two broad band ranges at the same time
(Marshall \cite{macho94}). These passbands did not correspond to standard filters but the blue one
had a mean wavelength similar to Johnson $V$. The camera covers a field of about 0.5 square
degree on the sky with a pixel scale of 0.62 arcseconds. The whole cluster could thus be observed
with one observation, and most of the cluster was on one CCD of the MACHO camera (there are four CCDs per filter band).
Observations were obtained once to twice a week. A few frames were lost due to
technical problems or bad seeing conditions.
Our monitoring came to an early end
after about five months when Mount Stromlo Observatory was destroyed by a bush fire.
Altogether we collected 15 useable frames over this time span. All observations were
done in queue observing mode.

For the determination of light curves and the detection of variables, only the blue
frames were used.  The light amplitude of long period variables is typically larger
in the blue than in the red (e.g. Fox \cite{fox82}) making the detection of variables
and the determination of periods easier.  Furthermore, many bright stars on the red frame were
over-exposed and thus not useable.  The blue detector had an area of bad pixels 
as well as a few scattered dead pixels.  Consequently, due to small positional shifts between
the different observing nights some stars could not be measured on all 15 frames. 

\subsection{CTIO data}

From August 2003 to January 2004, we continued the monitoring with ANDICAM at CTIO's 1.3m telescope operated
by the SMARTS consortium. The size of the CCD field is 6x6 arcminutes and the pixel scale is
0.37 arcsec/pixel. A description of the camera is given in DePoy et al.\,(\cite{andi03}). Due to the
smaller FOV we had to make a mosaic of images to cover most of the cluster.  Even then, we were limited
to the central part, and some of the outer areas covered by the Mount Stromlo data could not be observed.
Observations were done in $V$ and $I_{\rm c}$.  Observations were scheduled roughly once per week
giving a total of 16 useable frames.  As for the Mount Stromlo data,
observations were done in queue observing mode.

\subsection{Other data}

Six observations of the cluster in $V$ and $I_{\rm c}$ were taken in service mode with the WFI at the ESO 2.2m.
These observations were obtained between June and September 2002.
Four additional observations taken in July/August 2003 were kindly provided by Laszlo Kiss and collaborators. These observations
in $V$ and $I_{\rm c}$ were obtained with the Siding Spring Observatory 40-inch telescope, helping to
reduce the gap between the Mount Stromlo and the CTIO data set.

For a different project (Lebzelter et al.\,\cite{LWHJF05}), we did a short photometric monitoring of some
parts of 47 Tuc with ANDICAM at CTIO's 1.3m telescope in March to May 2002. These data were obtained in the $V$-filter with the old
ANDICAM CCD which had severe limitations (only half of the chip was useable).  These data were used only in rare cases.

\section{Data reduction}

For the Mount Stromlo and the CTIO data, flatfield and bias correction was done as part of the standard
data pipeline. The 40inch data were reduced applying standard data reduction with MIDAS, while the WFI data were reduced with the corresponding
IRAF package. For the detection of
variables and the measurement of the light curves on the Mount Stromlo and CTIO frames we used the image subtraction code ISIS 2.1 by
Alard (\cite{alard00}). First, the two data sets were analyzed separately. 
The reference flux of the identified variables required to produce light
curves from the image subtraction measurements was derived using the PSF
fitting software written by Ch.\,Alard for the
DENIS project. A description of this code can be found in Schuller et
al.\,(\cite{Schuller03}).
For the photometric
calibration of the CTIO data we used standard stars from Landolt's field RU\,149 (Landolt \cite{Landolt92}).
Photometric accuracy of the resulting light curves could be
analyzed by comparing measurements of variables in overlapping parts of the CTIO mosaic. Typical deviations
were of the order of 0.007\,mag.
We then used about 40 non-variable cluster stars ranging in $V-I$ between 0.9 and 2.1 to link the Mount Stromlo
measurements to those from CTIO.  Typical errors in the Mount Stromlo magnitudes were of the order of 0.013\,mag.

   \begin{figure*}
   \centering
   \resizebox{\hsize}{!}{\includegraphics{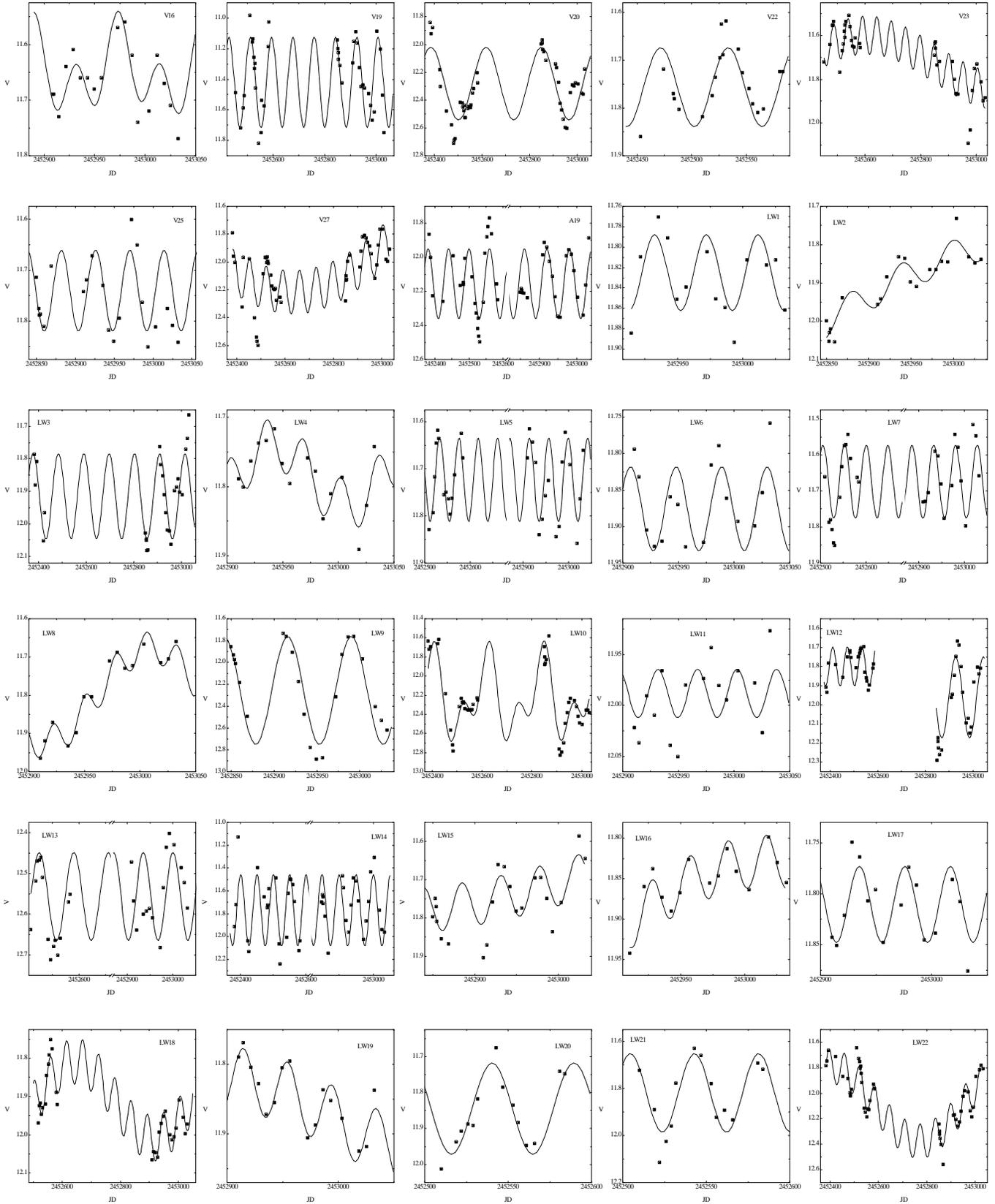}}
   \caption{Light curves of the variables of our survey, except those stars already discussed in Lebzelter
   et al. (\cite{LWHJF05}). Fourier fits with up to two periods are shown. Note that both axis scales vary from star to star.  For A19,
   LW5, LW7, LW13 and LW14, a break has been introduced into the time axis for a better presentation of the light change.  Errors on 
   data points before JD2452700 are estimated to be 0.013 mag. (Mount Stromlo data) while after this date the errors are 
   estimated to be 0.007 mag. (CTIO data).}
   \label{lctime}
   \end{figure*}

For various reasons (field of view, pixel scale, damaged parts of the CCD, depth of the observation)
not all variables could be measured on both the CTIO and the Mount Stromlo data. In these cases only one
of the data sets could be used for the analysis.

Our aim was to detect and measure long period variables on the upper part of the giant branch only. Thus
we selected for analysis only those variables that varied on time scales of more than
approximately 30 days with a total light amplitude of at least 0.1\,mag in $V$ (or the blue MACHO pass band).
For stars with shorter periods, our sampling was not sufficient to estimate a useful period. We searched
for periods in the selected light curves using the Fourier analysis code Period98 (Sperl \cite{sperl98}).
A maximum of two periods was derived for each star. First, the two data sets from Mount Stromlo and CTIO were
analyzed separately. The observations from ESO and SSO, measured using PSF fitting, were inserted in
the combined light curves. Then a period search was done on the whole light curve. Generally, the agreement
between the periods derived from these three data sets agreed very well. Naturally, stars with periods exceeding
the length of one of the data sets or with high irregularity in their light curve lead to deviating results.
The uncertainties of the periods calculated from a least square fit to the complete data sets are in most
cases less than a few percent. Of course the semiregular nature of most of the stars in our sample means that
the derived period is representative only for the current variability behaviour.
For the long period cases, we used only the period derived from the combined light curve. Some stars were
classified as irregular if no consistent period could be found. These stars will not be discussed further in
this paper.

\section{The variables}\label{var_section}

Periods have previously been published in the literature for the LPVs V1--V8, V11, V13, V18 and
V21 (we use the nomenclature of Clement et al.\,\cite{clement01}).  Light curves for these stars
derived from the current observations are given in Lebzelter et al.\,(\cite{LWHJF05}), along with some discussion of the origin
and reliability of the published periods.  These periods are given in column 7 of Table\,\ref{variables}.
Beside these stars, a few more red giants have
been identified as variables before, namely V15, V16, V17, V19, V20, V22, V23, V25, V27, V28 and A19. For some of these
(V15 to V17, V25 and V28) periods were given by Fox (\cite{fox82}).  A19 was reported variable by Lloyd-Evans 
(\cite{LE74}),
but no period was given. Later variability surveys, like those of Kaluzny et al.\,(\cite{kaluzny98}) and Weldrake et al.
(\cite{weldrake04}), have focused on binaries and RR\,Lyr stars and thus typically were not suitable for determining
periods of the stars on the upper giant branch
due to overexposure. The long period variables they report are most likely members of the Small Magellanic Cloud.

Table\,\ref{variables} lists the variables detected or characterized in the course of this study. Our sample includes 22
variables detected for the first time, and we give the first period determinations for six further stars previously
known to be variable.   All these periods are listed in column 6 of Table\,\ref{variables}.  We do not
include in Table\,\ref{variables} any previously known variables that we were not able to monitor (because
they fell on CCD defects or outside our field of view).  As nomenclature for new variables, we use LWxx with the numbering going from east to west.
Near infrared $J$ and $K$ magnitudes were extracted from the 2MASS database
and they are listed in columns 4 and 5.  The 2MASS coordinates are given in columns 2 and 3.

\begin{table*}
\caption{The LPVs of 47\,Tuc.} 
\begin{tabular}{lcccccccccc}
\hline 
Name & $\alpha$ (2000) & $\delta$ (2000) & $V$ & $V-I_{\rm c}$ & $J$ & $K$ & P [d] & P$_{literature}$ & Remark\\
\hline 
\noalign{\smallskip}
V1   & 00 24 12.4  & $-$72 06 39 & 13.15 &  3.86 & 7.45 & 6.21 & 221 & 212 & 1\\
V2   & 00 24 18.4  & $-$72 07 59 &   -    &   -    & 7.52 & 6.29 & 203 & 203 & 1,2\\
V3   & 00 25 15.9  & $-$72 03 54 & 12.63  &  3.54  & 7.49 & 6.27 & 192 & 192 & 1   \\
V4   & 00 24 00.3  & $-$72 07 26 & 12.34 &  2.62 & 7.87 & 6.69 & 165 & 82, 165 & 1,3 \\
V5   & 00 25 03.7  & $-$72 09 31 & 11.80  &  1.99  & 8.65 & 7.47 & 50 & 60 & 1,4  \\
V6   & 00 24 25.5  & $-$72 06 30 & 11.74 &  1.94 & 8.54 & 7.43 & 48 & 48 & 1  \\
V7   & 00 25 20.6  & $-$72 06 40 & 11.83  &  2.30  & 8.18 & 6.97 & 52 & 52 & 1,4    \\
V8   & 00 24 08.3  & $-$72 03 54 & 12.01 &  2.39 & 7.94 & 6.70 & 155 & 155 & 1    \\
V11  & 00 25 09.0  & $-$72 02 17 & 12.03  &  2.70  & 7.91 & 6.71 & 160: & 52, 100 & 4,5\\
V13  & 00 22 58.3  & $-$72 06 56 & 12.36  &  2.23  & 8.79 & 7.70 & 40 + long & 40 & 1,4,8\\
V16  & 00 25 23.2  & $-$72 11 05 & 11.65  &  2.0   & 8.35 & 7.23 & 41, 88: & 45 & \\
V18  & 00 25 09.2  & $-$72 02 39 & 11.67  &  1.95  & 8.59 & 7.47 & 83:  & & 1,4\\
V19  & 00 24 14.8  & $-$72 04 44 & 11.46 &  1.70 & 8.79 & 7.61 & 83 & & \\
V20  & 00 24 14.5  & $-$72 05 09 & 12.30   &  2.53 & 8.08 & 6.94 & 232: & & 11\\
V21  & 00 23 50.1  & $-$72 05 50 & 12.41 &  2.85 & 8.07 & 6.78 & 76: + long & & 1   \\
V22  & 00 24 08.9  & $-$72 03 00 & 11.80 &  1.99 & 8.38 & 7.18 & 62 & & 8 \\
V23  & 00 24 29.5  & $-$72 09 08 & 11.77 &  2.08 & 8.68 & 7.53 & 52 + long & & \\
V25  & 00 23 58.9  & $-$72 02 35 & 11.96 &  2.50 & 8.20 & 7.03 & 44 & 42 & 9\\
V27  & 00 24 15.2  & $-$72 04 36 & 12.11  &  2.52  & 7.97 & 6.77 & 69 + long & & \\
A19  & 00 24 21.8  & $-$72 04 13 & 12.18 &  2.57 & 8.08 & 6.79 & 60 & & \\
LW1  & 00 23 22.3  & $-$72 05 40 & 11.83 &  2.08 & 8.39 & 7.19 & 39 & & 9\\ 
LW2  & 00 23 29.2  & $-$72 06 20 & 11.90 &  2.23 & 8.26 & 7.05 & 60 + long & & 9\\ 
LW3  & 00 23 47.4  & $-$72 06 53 & 11.92 &  2.35 & 8.13 & 6.89 & 107 & & \\
LW4  & 00 23 51.3  & $-$72 03 49 & 11.80   &  1.98 & 8.56 & 7.37 & 32 + long: & & 9\\
LW5  & 00 23 53.2  & $-$72 04 16 & 11.74 &  1.96 & 8.53 & 7.32 & 40 & & \\ 
LW6  & 00 23 54.7  & $-$72 03 39 & 11.85  &  2.00 & 8.51 & 7.30 & 41 & & 9\\ 
LW7  & 00 23 56.9  & $-$72 05 33 & 11.66 &  1.86 & 8.26 & 7.18 & 42 & & 10\\ 
LW8  & 00 23 57.7  & $-$72 05 30 & 11.82 &  1.88 & 8.10 & 7.12 & 27: + long & & 9\\ 
LW9  & 00 23 58.2  & $-$72 05 49 & 12.34 &  2.77 & 7.97 & 6.74 & 74 & & 9\\ 
LW10 & 00 24 02.6  & $-$72 05 07 & 12.22 &   -    & 7.57 & 6.40 & 110:, 221: & & 2,10,11\\ 
LW11 & 00 24 03.2  & $-$72 04 51 & 11.95 &  1.79 & 8.47 & 7.40 & 36: & & 11\\ 
LW12 & 00 24 04.0  & $-$72 05 10 & 11.96 &  2.35 & 8.12 & 6.89 & 61, 116 & & 11\\ 
LW13 & 00 24 07.9: & $-$72 04 32:& 12.56 &  2.93 & 7.44: & 6.25: & 65: & & 6,11\\ 
LW14 & 00 24 09.4: & $-$72 04 49:& 11.71 &  1.99 & 8.42 & 7.39 & 50 & & 7,10\\ 
LW15 & 00 24 11.2  & $-$72 05 09 & 11.76 &  2.05 & 8.32 & 7.13 & 46 + long & & 9\\ 
LW16 & 00 24 13.6  & $-$72 04 52 & 11.88 &  1.89 & 8.39 & 7.32 & 29 + long & & 9 \\ 
LW17 & 00 24 16.3  & $-$72 01 31 & 11.81  &  1.99 & 8.43 & 7.20 & 41 & & 9\\ 
LW18 & 00 24 20.5  & $-$72 04 50 & 11.91  &  2.33 & 7.95 & 6.77 & 65 + long & & \\ 
LW19 & 00 24 23.2  & $-$72 04 23 & 11.85 &  2.17 & 8.38 & 7.16 & 40 + long & & 9,10 \\ 
LW20 & 00 24 52.1  & $-$71 56 11 &   -    &   -    & 8.31 & 7.11 & 49 & & 2,8\\ 
LW21 & 00 25 23.2  & $-$72 11 05 &   -    &   -    & 8.44 & 7.28 & 38 & & 2,8\\ 
LW22 & 00 25 30.1  & $-$72 04 32 & 12.06 &  2.48 & 8.16 & 6.93 & 63 + long & & \\ 
\noalign{\smallskip}
\hline
\noalign{\smallskip}
\end{tabular}

Notes: Column 2 and 3 give coordinates from 2MASS. Column 4 is the mean V brightness.
J and K colours are from 2MASS converted to the AAO system, if not stated otherwise. 
If there is an indication for
a long secondary period exceeding the time span of the monitoring, the star is
marked with '+long'. Uncertain values are marked with a colon.\\
Remarks: (1) see Lebzelter et al. (\cite{LWHJF05}); (2) no $V$ or $V-I_{\rm c}$, not in the CMD (Fig.\,\ref{lctime}); 
(3) currently only the longer period is
visible; (4) photometry in the CMD from Fox (\cite{fox82}) -- see text; (5) current periodicity
unclear, see Lebzelter et al. (\cite{LWHJF05}) for details; (6) close neighbour on 2MASS images, 
NIR fluxes and position possibly wrong; (7) not in 2MASS point source catalogue, 
coordinates calculated from our images, $K$ magnitude from
Origlia et al. (\cite{origlia02}); (8) only on Mount Stromlo frames; (9) no Mount Stromlo data; 
(10) stars with near infrared excess detected by Origlia et al. (\cite{origlia02}); 
(11) see text in Section\,\ref{var_section} for comments on the periods.
\label{variables}
\end{table*}

\begin{figure*}
   \centering
   \resizebox{\hsize}{!}{\includegraphics{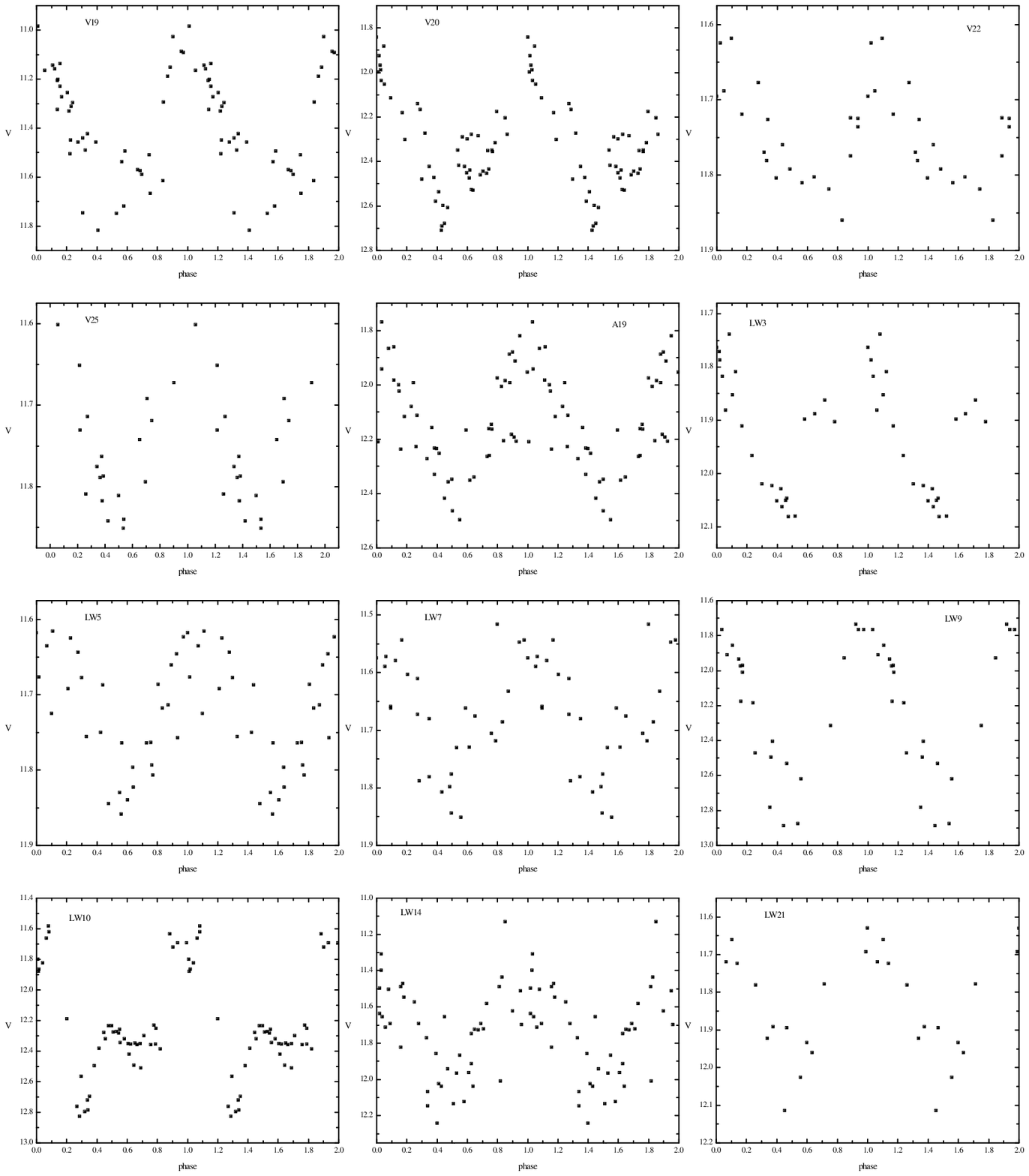}}
   \caption{Light curves versus phases for selected, monoperiodic variables from Table\,\ref{variables}. The period
   given in the table was used to calculate the phase.}
   \label{lcphase}
   \end{figure*}

Light variation versus time are presented in Figure\,\ref{lctime}, together with the fits from Fourier analysis. 
For selected variables with a reliable period determination and no long secondary period 
we show the light curve plotted against phase in Figure\,\ref{lcphase}. In Fig.\,\ref{lctime}
obvious deviations of the data from the fit illustrate the semiregular nature of many stars of the sample.
It can be seen
that quite a large fraction of variables shows a long secondary period lasting several hundred to several thousand days.
As these periods exceed the length of the time span monitored, we could not determine their length accurately and thus we do not
give them in Table\,\ref{variables}. They are included in the fits for illustrative purposes only. 
The formal ratio between the long and the short period in the plot ranges between 4 (LW4) and 100 (LW15). 
In most cases
the second period is 10 or more times longer than the shorter period. 

The Fourier fits to the stars V20, LW10, LW11, LW12 and LW13 deserve special comment. 
V20 and LW10 both appear to have periods of about 220-230 days but, because of the time gap in our observations,
these periods are quite uncertain.
Figs.\,\ref{lctime} and \ref{lcphase} suggest that these two stars have light curves with bumps during rising light,
a feature found in the light curves of several miras in the solar
neighbourhood. For LW10 the semi-period shows up clearly
in the Fourier analysis although this is not the case for V20.
The fit of LW11 is rather bad and the period determined is very uncertain.  In LW12,
the period and amplitude of the light curve clearly changed between the Mount Stromlo and CTIO observations
and we use separate Fourier fits to the two sets of observations.  The Mount Stromlo data set gave a period of
61\,d while the CTIO data give 116\,d. The fit to LW13 is poor, and the plotted fit curves include a
phase change of 20 days between the Mount Stromlo and CTIO observations.  Finally, we note that the 2MASS observations
could not separate LW13 from its close companion.

Multiperiodicity is a well known phenomenon among red variables in
the solar neighbourhood (e.g. Percy et al.\,\cite{percy03}, Kiss et al.\,\cite{kiss99}).
Long secondary periods are well known among long period variables, 
but a physical explanation for them has not yet been found
(Wood, Olivier \& Kawaler \cite{wok04}).

\begin{figure}
   \centering
   \resizebox{\hsize}{!}{\includegraphics{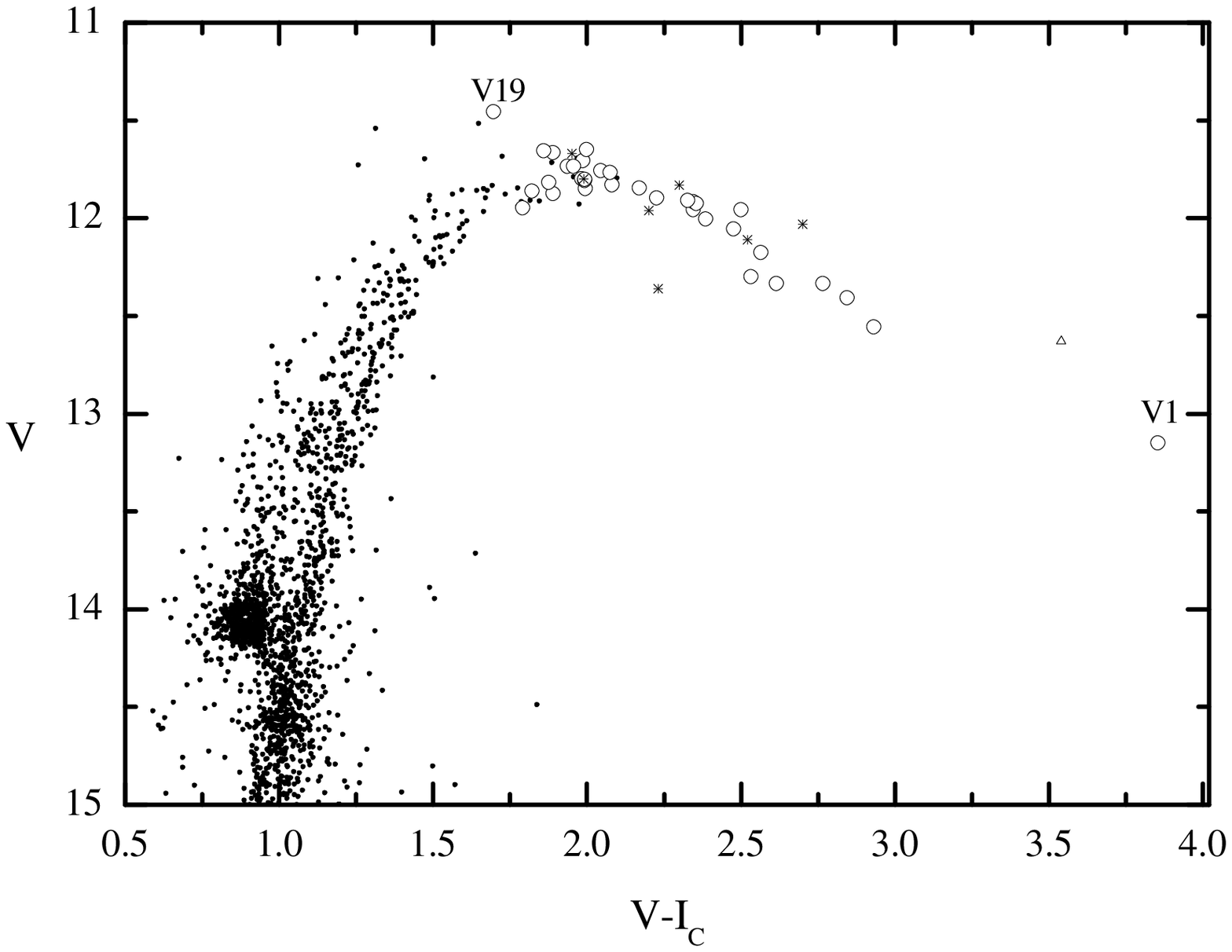}}
   \caption{Colour-magnitude diagram of 47\,Tuc. Variable stars are indicated by open circles. Asterisks mark 
   stars taken from Fox (\cite{fox82}) while the triangle is V3 using data from Eggen (\cite{eggen75}). 
   For all variables mean magnitudes and colours were used.}
   \label{cmd}
   \end{figure}

In Figure\,\ref{cmd} we show the position of the variables in the
colour-magnitude diagram using magnitudes from the CTIO images or, for a few
stars, an alternative source given below.  Only the giant branches and the
horizontal branch are shown.  Beside the variables listed in
Table\,\ref{variables}, we also marked stars detected as variables but slightly
below our amplitude cutoff criterion and stars with large irregularities in the
light curve so that a period could not be determined.  All $V$ and $V-I_{\rm
c}$ values of the variables are mean values derived from our light curves.  For
V5, V7, V11, V13 and V18 no CTIO data have been obtained.  Instead, we give the
values from Fox (\cite{fox82}).  V3 was also not observed at CTIO, but $V$ and
$I_K$ light curves were obtained by Eggen\,(\cite{eggen75}) and we use the mean
magnitudes from his observations, converting $V-I_K$ values to $V-I_{\rm c}$
using the transform in Bessell\,(\cite{bessell79}).  V2 and LW10 are missing
because the stars were saturated or unmeasurable on some of our $I$ frames.

Figure\,\ref{cmd} illustrates that all variables in our sample, with the
possible exception of V19, are almost certainly on the upper giant branch of
47\,Tuc (rather than in the SMC or Galactic halo).  V19 is located above and to
the left of the 47\,Tuc AGB. There is no indication for a long secondary period
in this star (see Fig.\,\ref{lctime}) which may lead to a non-representative
measurement\footnote{Long secondary periods may be responsible for some of the
scatter along the giant branch since the mean brightness is calculated over
only part of the light cycle corresponding to these periods.}. At the blue end
of the giant branch where V19 is found, its amplitude of more than 0.8\,mag is
rather untypical.  Possible explanations may be that it has a very close blue
neighbour that was not separated from V19 in our images, or it may be a star in
the Galactic halo.

The most interesting aspect of Figure\,\ref{cmd} is that it clearly shows that
all stars on the giant branch of 47\,Tuc redder than $V-I_{\rm c}$\,$\approx$\,1.8
are variable.

Finally, we comment on the connection between LPV pulsation and mass loss in
47\,Tuc.  Origlia et al.\,(\cite{origlia02}) listed five stars in 47\,Tuc with
a high infrared excess, indicating that they are surrounded by some dust. All
five are located within our investigated field. Identification is not simple as
Origlia et al. provide only near infrared finding charts of low spatial
resolution. Three stars from their list can be identified unequivocally with
optical counterparts (stars number 1, 3 and 5 from Origlia et al.). All three
are variables, namely LW10, LW14 and LW19, respectively. The source number 4
from Origlia et al. is probably related to the variable star LW7. The fifth
star listed in Origlia et al. (their number 2) could not be identified with any
LPV.  It is the star with the weakest infrared excess in their list.  It is
surprising that the star (LW19) with the largest mass loss rate in Origlia's
list (number 5) is neither one of the brightest variables nor has it a long
period. A similar comment can be made about V18, the variable with the largest
infrared excess in a survey of 47\,Tuc by Ramdani \& Jorissen (\cite{rj01}, see
also Lebzelter et al.\,\cite{LWHJF05}).  There is obviously no strict
correlation between pulsation period and mass loss rate.

\section{The $K$-log$P$ diagram}

\begin{figure}
\centering
\resizebox{\hsize}{!}{\includegraphics{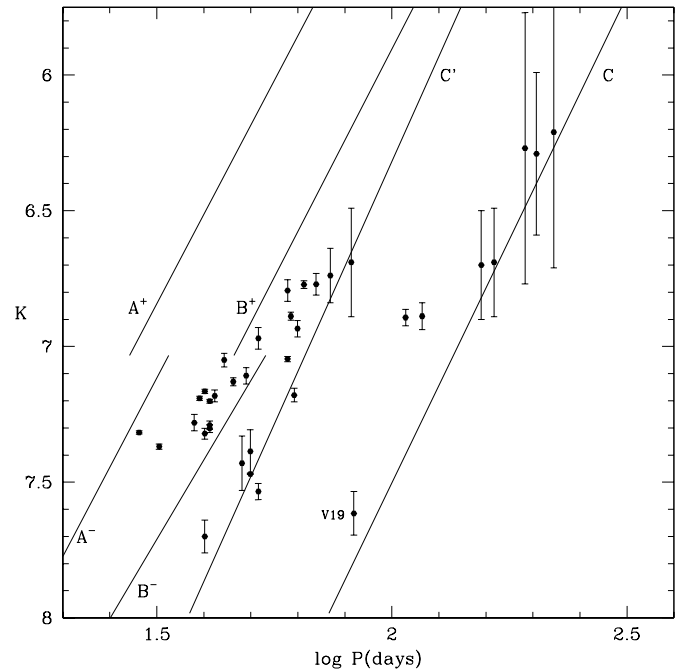}}
\caption{The $K$-$\log P$ diagram for the LPVs in 47\,Tuc. The error bars are scaled $V$ light
amplitudes.  Only variables with reasonably well-determined periods (periods without a following
colon in Table\,\ref{variables}) are shown.  
The lines show the sequences of LPVs in the LMC as given by Ita et al. (\cite{ita04}). Sequence C
is thought to be linked to fundamental mode pulsation, the other sequences to first to third
overtone pulsation.}
\label{klp_obs}
\end{figure}

Given the large number of LPVs now known in 47 Tuc, we are now in a position to
examine the PL relations in the cluster.  We do this in the form of the
$K$-$\log P$ relation, which is shown in Figure\,\ref{klp_obs}.  Generally, we
do not have $K$ light curves for the LPVs in 47 Tuc so we can not readily
calculate mean $K$ magnitudes.  However, a small number of the variables have
been observed many times (see Fox \cite{fox82}, Menzies \& Whitelock
\cite{mw85} and Frogel, Persson \& Cohen \cite{fpc81}) and good estimates of
the mean $K$ magnitude and $K$ amplitude can be obtained.  For each star
plotted on Figure\,\ref{klp_obs}, the $K$ magnitude is the mean of the maximum
and minimum observed $K$ magnitude.  For stars without multiple observations,
the 2MASS $K$ magnitude has been used.  All $J$ and $K$ magnitudes have been converted
to the AAO system using the conversions in Allen \& Cragg (\cite{ac83}) and
Carpenter (\cite{car01}): these are the values shown in Table~\ref{variables}.
A comparison of the $K$ amplitude with the $V$ light amplitude obtained in the
present study shows that the $K$ amplitude is approximately 20\% of the $V$
amplitude.  The error bars in Figure\,\ref{klp_obs} have a full length of 20\%
of the full visual amplitude of the pulsation mode associated with each point.
Hence, they should represent the $K$ amplitude of the variables.

Also shown in Figure\,\ref{klp_obs} are the sequences of LPVs in the LMC as given
by Ita et al. (\cite{ita04}).  The magnitudes have been adjusted to bring them
from an LMC distance modulus of 18.55 to a 47\,Tuc distance modulus of 13.50.
The large-amplitude Mira variables fall close to sequence C, while the bulk of
the smaller amplitude variables fall near, but not necessarily on, sequences
B$^{+}$, B$^{-}$ and C$^{\prime}$.  The lack of variables on sequences A$^{+}$
and A$^{-}$ may be real or it may be that these stars have amplitudes that are
below our detection limit of $\sim$0.1 mag.

It is notable in Figure\,\ref{klp_obs} that the 47\,Tuc LPVs do not appear to fall
exactly on the LMC sequences.  For example, the majority of the smaller amplitude
variables brighter than $K$=7 fall between sequences B$^{+}$ and C$^{\prime}$
while fainter than $K$=7, most variables fall between sequences A$^{-}$ and
B$^{-}$ with a small number appearing to fall on sequence C$^{\prime}$.  This
is probably a result of the different masses of the stars in the 47\,Tuc and LMC
samples.  We address this possibility in the next section.

If we neglect V19 which appears anomalously blue (see Sect.~\ref{var_section}),
stars seem to evolve up to $K$\,$\approx$\,6.7 in the 47\,Tuc equivalents of
sequences B and C$^{\prime}$, and to then switch to sequence C for the final
stage of evolution.  Using the results in Wood et al.\,(\cite{wm99}), this
switch seems to correspond to a transition in pulsation from a low order radial
overtone mode to the fundamental radial mode (compare also Lebzelter et al.\,\cite{LWHJF05}).

The tip of the RGB in the $K$-$\log P$ diagram occurs at $K=$7.1 (see Fig.\,\ref{klp_th}) at 
log\,$L$/L$_{\odot}$\,=3.355.
From Figs.\,\ref{klp_obs} and \ref{hrd} we can say that about half of the variables are more luminous than
the RGB tip and therefore must be AGB stars. The stars below the RGB tip could be
either on the AGB or the RGB.

\section{Pulsation models}
\subsection{Description of the models}
\begin{figure}
\centering
\resizebox{\hsize}{!}{\includegraphics{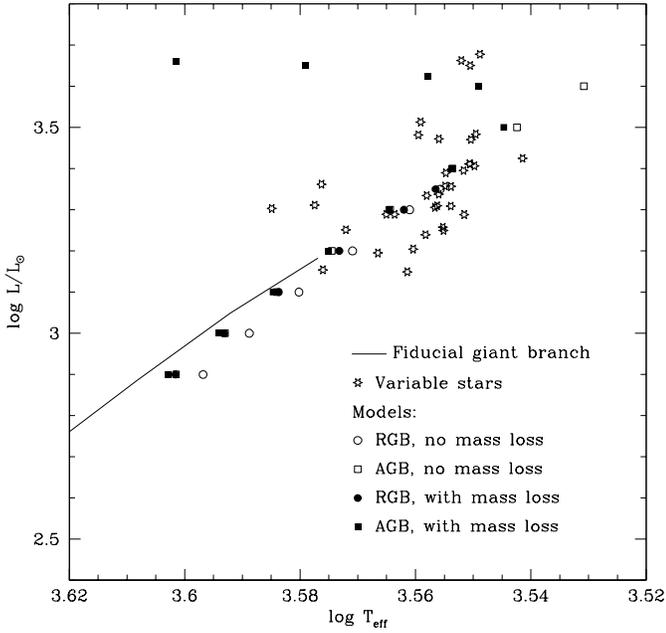}}
\caption{The tip of the 47\,Tuc giant branch in the HR-diagram.  $K$ and $J$-$K$ of the variables
were converted to log\,$L$/L$_{\odot}$ and $\log T_{\rm eff}$ using the transforms in Houdashelt et al. (\cite{hbs00}; \cite{hbsw00}).
The fiducial giant branch given by Hesser et al. (\cite{hes87}) is also shown,
with $V$ and $B$-$V$ also converted to log\,$L$/L$_{\odot}$ and $\log T_{\rm eff}$
using the transforms in Houdashelt et al.  The pulsation models are plotted using symbols identified
on the figure.}
\label{hrd}
\end{figure}

With a substantial set of LPVs now known in 47\,Tuc, we are in a position to
make theoretical models for the period-luminosity laws.  For our models, we
adopt the following parameters for 47\,Tuc (see the Introduction): distance
modulus $(m-M)_{V}$=13.5, reddening $E(B-V)$$=$0.024, helium mass fraction
Y=0.27, metal abundance Z=0.004 and main-sequence turnoff mass
0.9\,M$_{\odot}$.  Since we are computing pulsation periods, it is important to
get the radii (and hence $T_{\rm eff}$) of the models correct.  In order to
realize this condition, the mixing-length in the convection theory was set so
that the models coincided with the giant branch of 47\,Tuc.  Figure~\ref{hrd}
shows the models in the HR-diagram.  The giant branch of the models clearly
passes through the region of the giant branch occupied by the variables.

\begin{table*}
\caption{The pulsation models} 
\begin{tabular}{rccccrrrr}
\hline 
 $L$/L$_{\odot}$ &   $M$/M$_{\odot}$ & $M_{\rm c}$/M$_{\odot}$  & $\ell$/H$_p$  &  log\,T$_{\rm eff}$ &     P$_0$   &   P$_1$    &   P$_2$    &   P$_3$   \\
\hline
\multicolumn{9}{l} {RGB - no mass loss} \\
   794 &   0.9000 &   0.4002 &   1.80  &   3.5987  &    27.6  &   18.1  &   12.5  &    9.6 \\
  1000 &   0.9000 &   0.4136 &   1.80  &   3.5906  &    36.1  &   23.2  &   15.8  &   12.3 \\
  1259 &   0.9000 &   0.4274 &   1.80  &   3.5818  &    47.7  &   29.8  &   20.1  &   16.2 \\
  1585 &   0.9000 &   0.4417 &   1.80  &   3.5724  &    64.0  &   38.2  &   25.6  &   21.8 \\
  1995 &   0.9000 &   0.4565 &   1.80  &   3.5621  &    87.3  &   49.2  &   33.1  &   29.4 \\
  2239 &   0.9000 &   0.4640 &   1.80  &   3.5567  &   102.6  &   56.0  &   38.2  &   34.0 \\
\multicolumn{9}{l} {AGB - no mass loss} \\
   794 &   0.9000 &   0.4737 &   1.80  &   3.6034  &    26.5  &   17.3  &   11.9  &    9.2 \\
  1000 &   0.9000 &   0.4783 &   1.80  &   3.5948  &    34.8  &   22.2  &   15.2  &   11.8 \\
  1259 &   0.9000 &   0.4841 &   1.80  &   3.5857  &    46.2  &   28.6  &   19.3  &   15.7 \\
  1585 &   0.9000 &   0.4913 &   1.80  &   3.5760  &    62.1  &   36.9  &   24.7  &   21.2 \\
  1995 &   0.9000 &   0.5005 &   1.80  &   3.5656  &    84.8  &   47.6  &   32.1  &   28.5 \\
  2512 &   0.9000 &   0.5120 &   1.80  &   3.5546  &   117.7  &   61.6  &   43.0  &   37.8 \\
  3162 &   0.9000 &   0.5265 &   1.80  &   3.5427  &   166.8  &   79.9  &   59.8  &   48.9 \\
  3981 &   0.9000 &   0.5448 &   1.80  &   3.5304  &   242.8  &  104.2  &   84.9  &   63.4 \\
\multicolumn{9}{l} {RGB - with mass loss} \\
   794 &   0.8441 &   0.4002 &   1.90  &   3.6034  &    27.5  &   18.1  &   12.5  &    9.6 \\
  1000 &   0.8340 &   0.4136 &   1.90  &   3.5949  &    36.5  &   23.4  &   16.0  &   12.7 \\
  1259 &   0.8166 &   0.4274 &   1.90  &   3.5853  &    49.5  &   30.5  &   20.6  &   17.3 \\
  1585 &   0.7915 &   0.4417 &   1.90  &   3.5746  &    69.2  &   40.0  &   27.2  &   24.2 \\
  1995 &   0.7679 &   0.4565 &   1.90  &   3.5629  &    98.7  &   52.7  &   37.4  &   32.7 \\
  2239 &   0.7628 &   0.4640 &   1.90  &   3.5572  &   117.7  &   60.2  &   44.3  &   37.3 \\
\multicolumn{9}{l} {AGB - with mass loss} \\
   794 &   0.7371 &   0.4737 &   1.90  &   3.6045  &    29.9  &   19.1  &   13.1  &   10.5 \\
  1000 &   0.7344 &   0.4783 &   1.90  &   3.5956  &    39.9  &   24.6  &   16.7  &   14.1 \\
  1259 &   0.7301 &   0.4841 &   1.90  &   3.5862  &    53.9  &   31.7  &   21.5  &   19.0 \\
  1585 &   0.7244 &   0.4913 &   1.90  &   3.5762  &    74.0  &   40.9  &   28.5  &   25.3 \\
  1995 &   0.7118 &   0.5005 &   1.90  &   3.5654  &   104.6  &   53.2  &   39.4  &   32.9 \\
  2512 &   0.6903 &   0.5120 &   1.90  &   3.5539  &   154.2  &   70.0  &   56.8  &   42.8 \\
  3162 &   0.6567 &   0.5265 &   1.90  &   3.5441  &   237.1  &   93.6  &   81.9  &   57.0 \\
  3981 &   0.6057 &   0.5448 &   1.90  &   3.5478  &   321.2  &  123.2  &   99.9  &   74.1 \\
  4217 &   0.5904 &   0.5500 &   1.90  &   3.5571  &   320.6  &  124.6  &   96.1  &   73.9 \\
  4467 &   0.5743 &   0.5556 &   1.90  &   3.5789  &   291.6  &  113.7  &   81.3  &   66.7 \\
  4571 &   0.5675 &   0.5579 &   1.90  &   3.6012  &   266.7  &   98.6  &   67.0  &   57.3 \\
\hline 
\noalign{\smallskip}
\end{tabular}
\begin{flushleft}
Notes:  $\ell$/H$_p$ is the ratio of mixing-length to pressure scale height.  P$_0$,  P$_1$, P$_2$ and P$_3$\\ 
are the linear periods (in days) of the fundamental, 1st, 2nd and 3rd overtone modes.
\end{flushleft}
\label{puls_mods}
\end{table*}

The properties of the models are given in Table~\ref{puls_mods}.  The linear
non-adiabatic pulsation models were created with the pulsation code described
in Fox \& Wood (\cite{fw82}), updated to include interior opacities of
Iglesias \& Rogers (\cite{ir93}) and low temperature opacities of Alexander
\& Ferguson (\cite{af94}). This code uses a mixing length theory of
convection that explicitly treats variation of the convective velocity with
time.
The core mass $M_{\rm c}$ was obtained from the
$L$-$M_{\rm c}$ relation of Boothroyd \& Sackmann (\cite{bs88}).  The most
luminous RGB model corresponds to the RGB tip luminosity of the
0.9\,M$_{\odot}$, Z=0.004 tracks of Fagotto et al. (\cite{fbbc94}).

The models with mass loss were constructed according to the following
prescriptions.  Firstly, note that without mass loss the giant branch should
extend to very high luminosities well beyond
log\,$L$/L$_{\odot}$\,=\,10$^{4}$\,$L$/L$_{\odot}$.  Since the most luminous
stars on the 47\,Tuc giant branch have
$L$/L$_{\odot}$\,$\approx$\,4000$L$/L$_{\odot}$, it is assumed that mass loss
dissipates the hydrogen-rich envelope at about this luminosity.  Evolutionary
models with mass loss were constructed adopting a Reimers' mass loss law
(Reimers \cite{r75}), and the evolution rates on the the giant branch given by
the evolutionary tracks of Fagotto et al. (\cite{fbbc94}).  The Reimers mass
loss rate was multiplied by a factor $\eta$ which was adjusted so that the
stars left the AGB at $L$/L$_{\odot}$\,$\approx$\,4000 $L$/L$_{\odot}$ (a value
$\eta$ = 0.33 was required).  The resulting models are shown in the HR-diagram
Figure~\ref{hrd} and the masses of the models are given in
Table~\ref{puls_mods}.

\begin{figure}
\centering
\resizebox{\hsize}{!}{\includegraphics{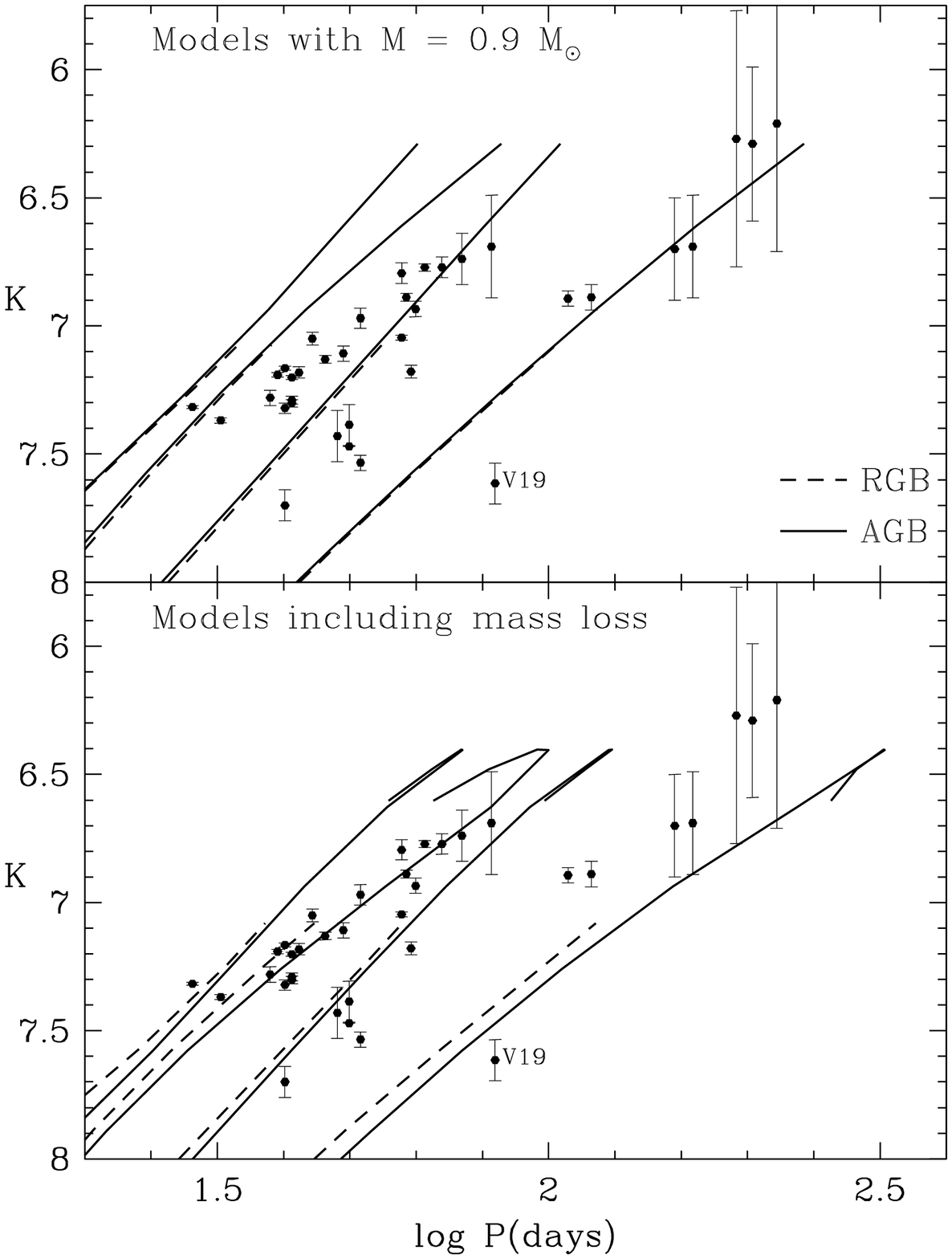}}
\caption{The $K$-$\log P$ diagram for models both with and without mass loss. The fundamental
mode and the first three overtones are shown.  The variables
with reasonably well-determined periods are also shown, as in Figure~\ref{klp_obs}.}
\label{klp_th}
\end{figure}

The $K$-$\log P$ diagram for models both with and without mass loss is compared
with the observed $K$-$\log P$ sequences in Figure~\ref{klp_th}. We now consider
the small and large amplitude sequences separately. 

\subsection{Small amplitude variables}
It is clear that 
the models without mass loss fail to reproduce the observed periods for the
smaller amplitude pulsators (the overtone pulsators, which have $\log P < 2$).
In fact the model sequences avoid the observed sequences.
Since the model calculations are for linear pulsation models, they should fit
the small amplitude variable sequences. 

A simple way to get the observed and model sequences to
agree would be to make the giant branch cooler at a given luminosity.  Since
the pulsation period for the overtones $P \propto R^{\frac{3}{2}}$ (Fox \&
Wood \cite{fw82}), and since $L=4\pi\sigma R^2 T_{\rm eff}^4$, at a given
luminosity $P \propto T_{\rm eff}^-3$.  In order to fit the theoretical overtone
sequences to the observed sequences, $\log P$ needs to increase by $\sim$0.15,
which means that $\log T_{\rm eff}$ needs to decrease by $\sim$0.05.  As can be
seen from Figure~\ref{hrd}, this is far greater than any likely uncertainty in
$\log T_{\rm eff}$.  We therefore conclude that models that have not undergone
considerable mass loss since the main-sequence can not explain the pulsation
periods of the overtone LPVs in 47\,Tuc.

In contrast to the models without mass loss, the overtone pulsation periods for
the stars that have undergone mass loss agree well with the observations.  The
periods of the mass loss models are longer than those of the models without
mass loss due to the lower stellar mass ($P \propto M^{-\frac{1}{2}}$), and the
slope of the $K$-$\log P$ relations is smaller due to the decrease in mass with
luminosity.  The decreased slope is particularly prominent for the second
overtone which moves from having a period close to that of the 3rd overtone at
low luminosities to a period close to that of the 1st overtone at high
luminosities.

\begin{figure}
\centering
\resizebox{\hsize}{!}{\includegraphics{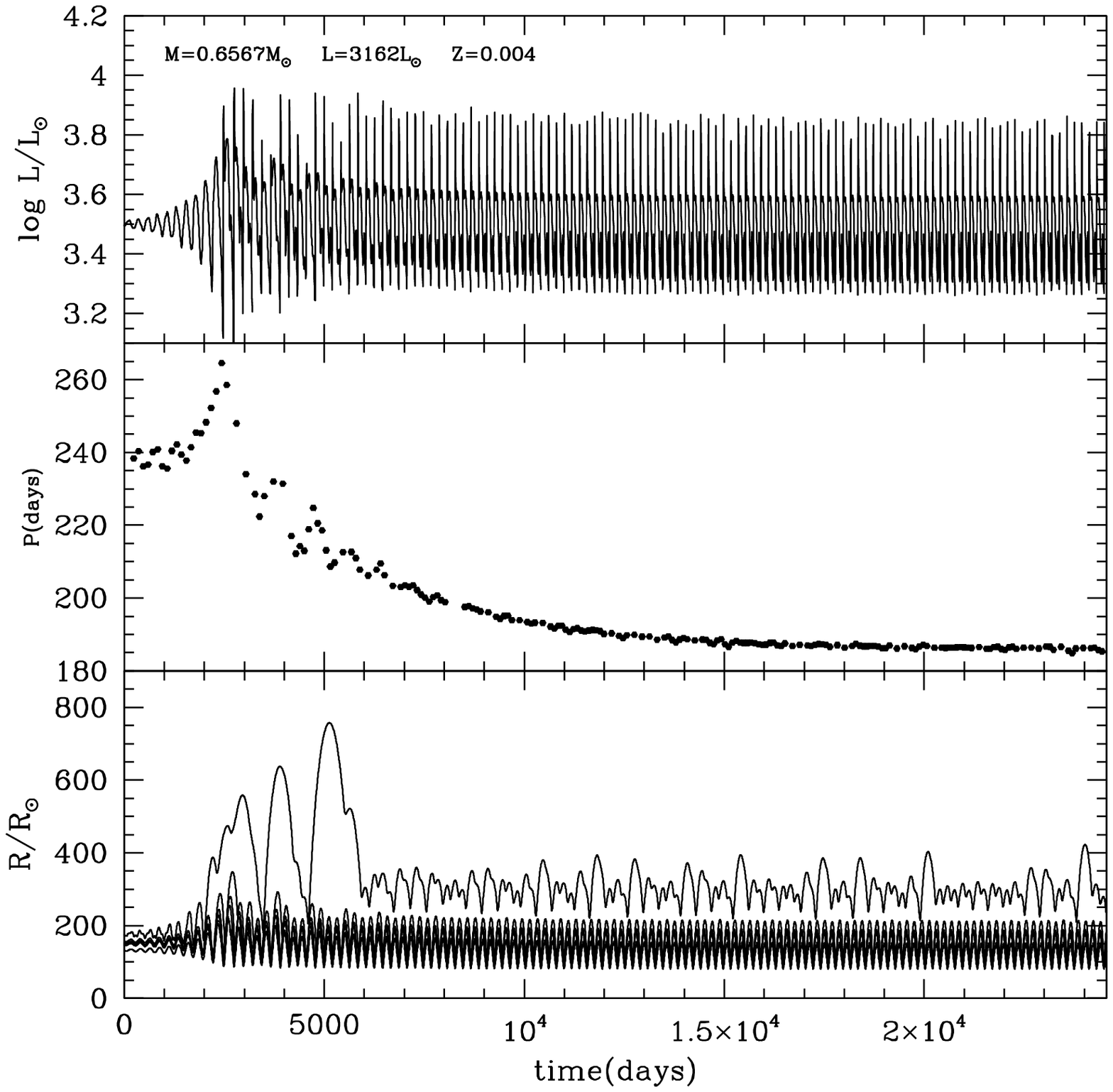}}
\caption{Luminosity $L$, pulsation period $P$ and radii of several mass points
plotted against time in a model with $M$ = 0.6567 M$_{\odot}$ and $L$ = 3162 L$_{\odot}$.
The linear pulsation period of this model is 240 days while the nonlinear pulsation
period after 67 years of relaxation is 185 days.
}
\label{p_change}
\end{figure}

\subsection{Large amplitude variables}
In contrast to the small amplitude variables, the large-amplitude variables
i.e.~the Miras are consistent with theoretical models without mass loss, while
the models with mass loss do not fit the observed $K$-$log P$ sequence. 
(We ignore V19 in this comparison since Figure~\ref{cmd}
suggests this star is a Galactic halo star rather than a 47\,Tuc star.)

We believe this contradiction can be explained by the nonlinear effects
in the pulsation of red giants with large amplitudes. 
Some preliminary nonlinear pulsation
calculations made for these models show that the full-amplitude pulsation
periods are considerably shorter than the linear periods due to a change in the
envelope structure associated with large amplitude pulsation (see also Ya'Ari
\& Tuchman \cite{yt96}).  An example of the change in period of one of these
models, perturbed with a low amplitude and allowed to reach limiting amplitude,
is shown in Figure~\ref{p_change}.  Thus the periods of {\em nonlinear}
fundamental mode models may be able to explain the periods of the Miras in
47\,Tuc, although this needs further exploration.  The nonlinear models will be
described elsewhere.

\begin{figure}
\centering
\resizebox{\hsize}{!}{\includegraphics{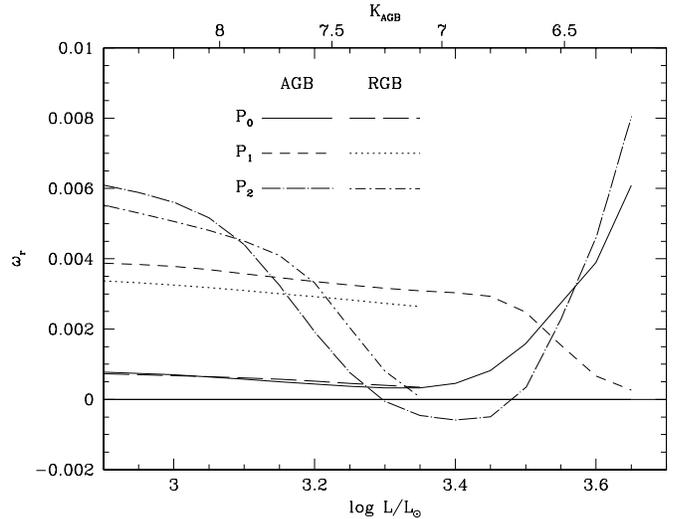}}
\caption{The growth rate $\omega_{\rm r}$ (days$^{-1}$ plotted against log $L$/L$_{\odot}$ for the
models in the mass loss sequence.  $\omega_{\rm r}$ is the real part of the eigenvalue $\omega$
where a time dependence exp($\omega t$) is assumed.  The $K$ magnitude for AGB stars is shown
on the top axis.
}
\label{gr}
\end{figure}

\subsection{Evolution of pulsation mode with luminosity}
Figure~\ref{klp_th} provides an indication of the pulsation evolution
of stars evolving up the giant branch.  Ignoring V19 (as noted above), it seems that stars
start pulsating at $K \sim 7.7$ in the 1st to 3rd overtone, then evolve
up to higher luminosities to $K \sim 6.9$ where the stars
transit to fundamental mode pulsation.  Further evolution
is in the fundamental mode, until mass loss terminates AGB evolution
at $K \sim 6.2$.  This is broadly consistent with what is expected from the linear
non-adiabatic growth rates of the models.  These are shown plotted against luminosity in Figure~\ref{gr}.
It should be remembered that the theoretical growth rates for these highly convective stars
are very uncertain and should be regarded as indicative only.  The general behaviour shown in 
Figure~\ref{gr} is that, at low luminosities the 2nd overtone
has the highest growth rate, at intermediate luminosities the 1st overtone has the
highest growth rate, and at high luminosities the fundamental mode grows most rapidly.
The 2nd overtone also has a high growth rate at high luminosities but it is 
likely to be overwhelmed by the fundamental mode in the nonlinear case.
This suggests an evolution from 2nd to 1st overtone and then to fundamental mode.
This is similar to the observed situation, although the overtones co-exist
rather than following a distinct progression with luminosity.

\section{Summary and Conclusions}

We have identified 22 new variable red giants in 47\,Tuc and determined periods
for another 8 previously known variables.  All red giants redder than
$V$-$I_{\rm c}$\,=\,1.8 are variable at the limits of our detection threshold,
which corresponds to $\delta V$\,$\approx$\,0.1 mag.  This colour limit
corresponds to a luminosity log\,$L$/L$_{\odot}$=3.15 and it is considerably
below the tip of the RGB at log\,$L$/L$_{\odot}$=3.35.  In the $K$-log$P$
diagram, the 47\,Tuc variables do not closely follow the ridge lines of PL
relations seen for LPVs in the MCs, indicating that the PL relations are mass
dependent.

Linear non-adiabatic modelling was used to try to reproduce the observed PL
relations, especially for the low amplitude pulsators where linear calculations
should be appropriate.  It was shown that models without mass loss can not
reproduce the observed PL laws for the low amplitude pulsators.  Models that
lose sufficient mass to terminate AGB evolution near $L \sim 4000
L$/L$_{\odot}$ do reproduce the observed PL relations for low amplitude
variables.  This is the first time that measurements of the masses of stars on
the AGB have shown that mass loss of the order of 0.3 M$_{\odot}$ occurs along
the RGB and AGB.

The linear pulsation periods do not agree well with the observed periods of the
large amplitude Mira variables, which pulsate in the fundamental mode.  The
solution to this problem appears to be that the nonlinear pulsation periods in
these low mass stars are considerably shorter than the linear pulsation periods
due to a rearrangement of stellar structure caused by the pulsation.  Although
such effects have been seen in pulsation models before, the 47\,Tuc stars
studied here provide the first observational evidence for this effect.

The observations show that stars evolve up the RGB and first part of the AGB
pulsating in low order overtone modes, then switch to fundamental mode at high
luminosities.  The linear non-adiabatic growth rates of models suggest that
such behaviour should occur but the models at this stage are only indicative.
It is hoped that future improved models including the effect of turbulent
viscosity (e.g. Olivier \& Wood \cite{ow05}) will allow a reliable
determination of the growth rates and mode selection processes in red giant
stars, as well as an estimation of the effect of nonlinear pulsation on the
pulsation period.

\begin{acknowledgements}

We are indebted to Laszlo Kiss for kindly providing additional images for our
time series of 47 Tuc. We wish to thank Christophe Alard for support with the
image subtraction code ISIS and for providing his PSF fitting software.  We
thank Brian Schmidt for organizing the monitoring of 47\,Tuc at Mount Stromlo
Observatory, and the queue mode observers at ESO 2.2m and CTIO 1.3m. This
project obtained data via the NOAO share of the SMARTS consortium. TL has been
supported by the Austrian Academy of Science (APART programme).  PRW has been
partially supported by a grant from the Australian Research Council.  This
publication makes use of data products from the Two Micron All Sky Survey,
which is a joint project of the University of Massachusetts and the Infrared
Processing and Analysis Center/California Institute of Technology, funded by
the National Aeronautics and Space Administration and the National Science
Foundation.

\end{acknowledgements}


\begin{thebibliography}{}

\bibitem[2000]{alard00} Alard, C., 2000, A\&A Suppl., 144, 363
\bibitem[1994]{af94} Alexander, D.R., Ferguson, J.W., 1994, ApJ, 437, 879
\bibitem[2001]{ac83} Allen, D.A., Cragg, T.A., 1983, MNRAS, 203, 777
\bibitem[1992]{macho92} Alcock, C., Axelrod, T.S., Bennett, D.P., et al., 1992, in: {\it Robotic Telescopes in the
1990s}, ed. A.V. Fillippenko, ASP Conf. Ser. 34, p.193
\bibitem[1979]{bessell79} Bessell, M.S., 1979, PASP, 91, 589
\bibitem[1988]{bs88} Boothroyd, A,I., Sackmann, I.-J., 1988, ApJ, 328, 641
\bibitem[1997]{cg97} Caretta, E., Gratton, R.G., 1997, A\&AS, 121, 95
\bibitem[2001]{car01} Carpenter, J.M., 2001, AJ, 121, 2851
\bibitem[2001]{clement01} Clement, C.M., Muzzin, A., Dufton, Q., et al., 2001, AJ, 122, 2587
\bibitem[2003]{andi03} DePoy, D.L., Atwood, B., Belville, S.R., et al., 2003, SPIE 4841, 827
\bibitem[1975]{eggen75} Eggen, O.J., 1975, ApJ, 195, 661
\bibitem[1994]{fbbc94} Fagotto, F., Bressan, A., Bertelli, G., Chiosi, C., 1994, A\&AS, 105, 29
\bibitem[1982]{fox82} Fox, M.W., 1982, MNRAS, 199, 715
\bibitem[1982]{fw82} Fox, M.W., Wood, P.R., 1982, ApJ, 259, 198
\bibitem[2005]{fra05} Fraser, O.J., Hawley, S.L., Cook, K.H., Keller, S.C., 2005, AJ, 129, 768
\bibitem[1981]{fpc81} Frogel, J.A., Persson, S.E., Cohen, J.G., 1981, ApJ, 246, 842
\bibitem[2003]{gratton03} Gratton, R.G., Bragaglia, A., Carretta E., et al., 2003, A\&A, 408, 529
\bibitem[1987]{hes87} Hesser, J.E., Harris, W.E., Vandenberg, D.A., Allwright, J.W.B., Shott, P., Stetson, P.B., 1987, PASP, 99, 739
\bibitem[2000a]{hbs00} Houdashelt, M.L., Bell, R.A., Sweigart, A.V., 2000, AJ, 119, 1448
\bibitem[2000b]{hbsw00} Houdashelt, M.L., Bell, R.A., Sweigart, A.V., Wing, R.F., 2000, AJ, 119, 1424
\bibitem[1993]{ir93} Iglesias, C.A., Rogers, F.F., 1993, ApJ, 412, 572
\bibitem[2004]{ita04} Ita, Y., Tanab\'{e}, T, Matsunaga, N., Nakajima, Y., Nagashima, C., Nagayama, T., Kato, D., Kurita, M.,
  Nagata, T., Sato, S., Tamura, M., Nakaya, H., Nakada, Y., 2004, MNRAS, 347, 720
\bibitem[1998]{kaluzny98} Kalzuny, J., Kubiak, M., Szymanski, M., et al., 1998, A\&A Suppl., 128, 19
\bibitem[1999]{kiss99} Kiss, L.L., Szatmary, K., Cadmus, R.R., Mattei, J.A., 1999, A\&A, 346, 542
\bibitem[1992]{Landolt92} Landolt, A.U., 1992, AJ, 104, 340
\bibitem[2004]{lebz04} Lebzelter, T., Wood, P.R., Hinkle, K.H., Joyce, R.R., Fekel, F.C., 2004, Proceedings of IAU Coll. 193, 
  "Variable stars in the Local Group", ASP Conf.~Ser.~310, p.144
\bibitem[2005]{LWHJF05} Lebzelter, T., Wood, P.R., Hinkle, K.H., Joyce, R.R., Fekel, F.C., 2005, A\&A, 432, 207
\bibitem[1974]{LE74} Lloyd-Evans, T., 1974, MNRAS, 167, 393
\bibitem[1994]{macho94} Marshall, S., 1994, in: {\it Astronomy from wide-field imaging},
  IAU Symp.~161, eds.~H.T.MacGillivray et al., Kluwer, Dordrecht, p.67
\bibitem[1985]{mw85} Menzies, J.W., Whitelock, P.A., 1985, MNRAS, 212,783
\bibitem[2005]{ow05} Olivier, E.A., Wood, P.R., 2005, submitted
\bibitem[2002]{origlia02} Origlia, L., Ferraro, F.R., Fusi Pecci, F., Rood, R.T., 2002, ApJ, 571, 458
\bibitem[2003]{percy03} Percy, J.R. Bessla, G., Velocci, V., Henry, G.W., 2003, PASP, 115, 479
\bibitem[2001]{rj01} Ramdani, A., Jorissen, A., 2001, A\&A, 372, 85
\bibitem[1975]{r75} Reimers, D. 1975, in Problems in Stellar Atmospheres and Envelopes, eds. B.
Bascheck, W.H. Kegel and G. Traving (Springer: Berlin), p.229
\bibitem[2003]{Schuller03} Schuller, F., Ganesh, S., Messineo, M., et al., 2003, A\&A, 403, 955
\bibitem[2004] {sos04} Soszy\'{n}ski, I., Udalski, A., Kubiak, M., Szyma\'{n}ski, M., Pietrzy\'{n}ski, G., Zebru\'{n}, K.,
Szewczyk, O., Wyrzykowski, L., 2004, Acta Astr., 54, 129
\bibitem[1998]{sperl98} Sperl, M., 1998, Comm.\,Asteroseism., 111, 1
\bibitem[2004]{weldrake04} Weldrake, D.T.F., Sackett, P.D., Bridges, T.J., Freeman, K.C., 2004, AJ, 128, 736
\bibitem[1999]{wm99} Wood, P.R. and the MACHO Collaboration, 1999, in IAU Symposium 191, Asymptototic Giant
Branch Stars, eds. T. Le Bertre, A. L\`{e}bre \& C. Waelkens (San Francisco:
ASP), p. 151
\bibitem[2000]{wood00} Wood, P.R., 2000, PASA, 17, 18
\bibitem[2004]{wok04} Wood, P.R., Olivier, E.A., Kawaler, S.D., 2004, ApJ, 604, 800
\bibitem[1996]{yt96} Ya'Ari, A., Tuchman, Y., 1996, ApJ, 456, 350

\end{thebibliography}
\end{document}